\documentclass[aps,twocolumn]{revtex4}
\usepackage{graphicx}
\usepackage{epsfig}
\def \Tan {{\rm Tan}}
\def \m {m_j}
\def \r {{\bf r}_{12}}
\def \ds {\displaystyle}

\def \al {\alpha }

\def \hm {{\bf m}}

\def \TC {T_{\rm c}}

\def \vS {{\bf S}}

\def \seff {A_{\rm eff}}

\def \Re {{\rm Re}} 
\def \uI {\underline{I}}
\def \uG {\underline{G}}

\def \Jc {J_c}
\def \tJc {\tilde{J}_c}
\def \G0 {\Gamma^{(0)}}
\def \vJ {\vec J}
\def \vk {{\bf k}}
\def \vR {{\bf R}}
\begin{document}

\date{\today}
\title{Transition Temperature of a Magnetic Semiconductor with Angular Momentum $j$}
\author{Juana Moreno$^1$, Randy S. Fishman$^{2 }$, and Mark Jarrell$^{3}$}
\affiliation{$^1$Physics Department, University of North Dakota, Grand Forks, 
North Dakota 58202-7129}
\affiliation{$^{2}$Condensed Matter Sciences Division, Oak Ridge National Laboratory, 
Oak Ridge, Tennessee 37831-6032}
\affiliation{$^{3 }$Department of Physics, University of Cincinnati, Cincinnati, 
Ohio 45221-0011}

\begin{abstract}

We employ dynamical mean-field theory to identify the
materials properties that optimize $\TC$ for a generalized
double-exchange (DE) model.  We reach the surprising conclusion that $\TC $
achieves a maximum when the band angular momentum $j$ equals 3/2 and when the
masses in the $\m =\pm 1/2$ and $\pm 3/2$ sub-bands are equal.
However, we also find that $\TC $ is significantly reduced as 
the ratio of the masses decreases from one.
Consequently, the search for dilute magnetic semiconductors (DMS) 
materials with high $\TC $ should 
proceed on two fronts. In semiconductors with $p$ bands, such as the 
currently studied Mn-doped Ge and GaAs semiconductors,
$\TC $ may be optimized by tuning the band masses through strain
engineering or artificial nanostructures.  On the other hand, 
semiconductors with $s$ or $d$ bands
with nearly equal effective masses might prove to have higher $\TC $'s
than $p$-band materials with disparate effective masses.

\end{abstract}
\maketitle

The discovery of dilute-magnetic semiconductors (DMS) with current transition temperatures above 
170 K\cite{Ohno96,mun:89,MacDon05} initiated
an active search for the optimal material for spintronic device applications \cite{Zutic04}.
However, the ferromagnetic transition temperature of magnetic semiconductors 
involves many parameters that are difficult to control experimentally. 
To facilitate the search for new magnetic semiconductors with high transition temperatures, 
we evaluate $\TC $ for a double-exchange system with general angular momentum $j$ using 
dynamical mean-field theory (DMFT).
Surprisingly, $\TC $ is found to reach a maximum for $j=3/2$ and for equal light and heavy 
hole masses.  However, we also find that $\TC $ decreases significantly 
when the ratio of the masses is reduced.

The notion of using magnetic semiconductors in spintronic devices dates back to the 
1960's, when europium chalcogenides \cite{Mauger86} and chromium spinels \cite{Baltzer66} were
extensively studied. 
Before the appearance of the latest generation of III-V DMS grown by 
molecular beam epitaxy techniques \cite{Ohno96,mun:89},  
II-VI \cite{Furdyna88} and IV-VI \cite{Story86} DMS were developed by 
alloying non-magnetic semiconductors with magnetic ions. 
Improvement in growth techniques has pushed the $\TC$ of Ga$_{1-x}$Mn$_x$As
to values above 170 $K$ \cite{MacDon05}.
As we discuss later, the large value of  $\TC$
in delta-doped GaAs \cite{Nazmul03} might be associated with the reduced magnetic
frustration when the Mn ions are restricted to two-dimensional planes.

In our calculation we employ the DMFT, which was
formulated in the late 1980's by M\"uller-Hartmann \cite{muller89} and Metzner and 
Vollhardt \cite{met:89}.  It has since developed into one of the most powerful many-body techniques 
for studying electronic models such as the Hubbard \cite{fre:95,geo:96} and DE 
\cite{fur:95,mil:96,mich03,aus:01,fis:03,che:03} models.  
Since DMFT becomes exact in the dilute limit, it is a good starting 
point for studying DMS.
Recent work on DMS materials has used DMFT to study variants of the DE 
model \cite{cha:01,ara:04} with less than one local moment per site.  
A DE model with 
one local moment per site and large coupling constant $\Jc$ provides an upper limit to the
transition temperature for a system with exchange coupling between the local moments
and charge carriers.  Perhaps more importantly, the behavior of $n_h < x$ holes in the
impurity band of a DMS with $x < 1$ Mn atoms per site is very similar to that of 
a DE model with filling $p=n_h/x < 1$ \cite{ara:04}.  So as long as $\Jc $ is sufficiently
large to produce a well-defined impurity band, the qualitative results of this
model should be independent of the precise magnitude of $\Jc /W$ \cite{che:03}.  
As we shall see, a generalized DE model with one local moment per site and large $\Jc $ also 
has the distinct advantage that analytic results are possible for any angular momentum
$j$ of the charge carriers. 

In semiconductors like GaAs, the angular momentum of the holes is obtained from
the vector sum of the $s=1/2$ spin with the $l=1$ orbital angular momentum
of the $p$ bands.  The $j=3/2$ band lies highest in energy while the spin-orbit split 
$j=1/2$ band lies an energy $\Delta_{so}\approx 340$ meV below \cite{bla:82}.  Consequently, 
almost all of the holes in Mn-doped GaAs populate the $j=3/2$ band, which in turn is split 
by crystal fields \cite{lut:56} into a $\m =\pm 3/2$ sub-band with heavy holes and a $\m =\pm 1/2$ 
sub-band with light holes.  
More exotic semiconductors with ($l=2$) $d$ bands, such as the chalcogenides
\cite{Mauger86}, may contain carriers with total angular momentum $j=5/2$.   

The optimization of $\TC $ is performed in two stages.  First, we calculate $\TC $ 
for a DE model where the charge carriers have angular momentum $j$ but equal masses in 
all $2j+1$ sub-bands.  After concluding that the system with $j=3/2$ and a half-filled 
lower or upper band has the highest $\TC$, we examine the effect of different band 
masses in the $\m =\pm 1/2$ and $\pm 3/2$ sub-bands.

The Hamiltonian of a generalized DE model with carriers (holes or electrons) of 
angular momentum $j$ and equal masses in all sub-bands is given by
\begin{equation}
\label{ham}
H=\sum_{\vk }\epsilon_{\vk }c^{\dagger }_{\vk \al }c^{\, }_{\vk \al }
-\frac{\Jc }{N}\sum_{i,\vk ,\vk'} e^{i(\vk -\vk')\cdot \vR_i} \vS_i\cdot 
c^{\dagger }_{\vk' \al } \vJ_{\al \beta }c_{\vk \beta }, 
\end{equation}
where $c^{\dagger }_{\vk \al }$ and $c_{\vk \al }$ are the creation and destruction operators 
for an electron with angular-momentum component $\m = \al $ ($\al =-j,-j+1,\ldots ,j$) and momentum $\vk $,
$\vS_i=S\hm_i $ is the spin of the local moment (treated classically) at site $\vR_i$, and $\vJ_{\al \beta }$ are the 
$(2j+1)$-dimensional angular-momentum matrices.  Repeated spin indices are summed.  

Within DMFT, 
the local effective action $\seff (\hm )$ 
is governed by the bare Green's function $G_0$, 
which contains dynamical information about the hopping of electrons onto and off neighboring sites.
Because $\seff (\hm )$ is quadratic in the field variables, the full local 
Green's function $G(i\nu_n)_{\al \beta }$ may be readily solved by integration, 
with the result \cite{fur:95} 
$\uG(i\nu_n ) =\langle \bigl(\uG_0(i\nu_n)^{-1} +\tJc \underline{\vJ } \cdot \hm \bigr)^{-1}\rangle_{\hm }$,
where  $\nu_n=(2n+1)\pi T$ are  Matsubara frequencies and
underlined quantities are matrices in $(2j+1) \times (2j+1)$ spin space;
$\langle X(\hm )\rangle_{\hm } =\int d\Omega_{\hm }X(\hm )P(\hm) $ denotes 
an average over the orientations $\hm $ of the
local moment with a probability $P(\hm)$, and $\tJc =\Jc S$.  
This relation is solved for a semicircular density-of-states
with full bandwidth $W$ in terms of the electron filling $p$ ($p=1$ corresponds to one electron per 
site so that $0 \le p \le 2j+1$) and the local-moment order parameter
$M=\langle m_z\rangle_{\hm }$, which becomes nonzero below $\TC $.

As $\tJc $ increases, the $(2j+1)$-degenerate band splits into $2j+1$ sub-bands, each 
labeled by quantum number $\m $ and centered at energy $-\m \tJc $.
Due to the effect of electronic correlations, the full bandwidth of each sub-band 
is lowered from $W$ to $W'=W/\sqrt{2j+1}$.  So prior to taking the limit of large 
$\Jc $ for the $\m $ sub-band, we must rewrite the chemical potential as $\mu =-\m \tJc +\delta \mu $ 
where $\vert \delta \mu \vert \le W'/2$.  

Since matrix quantities like $\uG(i\nu_n)$, $\uG_0(i\nu_n)$, and the self-energy 
$\underline{\Sigma }(i\nu_n)= \uG_0(i\nu_n)^{-1}-\uG(i\nu_n)^{-1}$ have already been
averaged over the orientations of $\hm $, they are spin diagonal and may be expanded in 
powers of $\underline{J}_z$ as
$\underline{A}=A_0 \uI +A_1 \underline{J}_z +A_2\underline{J}_z^2+\ldots $
where $A_n$ is of order $M^n$.  
So to linear order in $M$, only the first two terms in this
expansion are retained and the bare inverse Green's function may be parametrized as 
$\uG_0(i\nu_n )^{-1} =(z_n-\m \tJc +R_n)\uI +Q_n\underline{J}_z$ where $z_n=i\nu_n+\delta \mu $.

For large $\Jc $ and to linear order in $M$, we solved self-consistent equations for
$R_n$ and $Q_n$. The Curie temperature satisfies the condition 
\begin{equation}
\label{tcs}
\sum_n \ds\frac{R_n^2}{R_n^2-j(j+1)W'^2/16\m^2}=1,
\end{equation}
while the filling  
$\displaystyle p=T \sum_n Tr \Re [\uG(i\nu_n)] +j-\m +\ds\frac{1}{2}$.
These two expressions generalize previous results \cite{fur:95,fis:03} 
for $j=1/2$. Quite naturally, a system without partially filled sub-bands ($p$ an integer)
has a vanishing Curie temperature because the carriers are unable to hop to 
neighboring sites without incurring an infinite cost in coupling energy.

For a given $j$, the largest $\TC $ always occurs in the sub-bands with $\m =\pm j$ 
because those holes or electrons are able to most effectively take advantage of the 
exchange coupling that mediates the ferromagnetism between the local moments.
For $j=1/2$, 3/2, 5/2, 7/2, and 9/2, the dependence of $\TC $ on filling  within the
$\m =\pm j$ sub-bands is plotted in Fig.~1(a).  As expected, $\TC $ is particle-hole symmetric 
within each sub-band and is the same for a system with $p$ electrons ($1-p$ holes) or $p$ holes 
($1-p$ electrons) per site.  So the largest Curie temperature, $\TC^{{\rm max}}$, 
is obtained when the lowest or highest sub-band is half-filled.

For a half-filled sub-band with $\m =\pm j$, we obtain an analytic expression for 
$\TC^{{\rm max}} $ by 
converting the Matsubara sum into an integral (assuming that $\TC^{{\rm max}} /W$ is small), 
with the result 
\begin{equation}
\ds\frac{\TC^{{\rm max}}}{W}=\frac{1}{4\pi \sqrt{2j+1}}\Biggl\{
1-\ds\frac{1}{2\sqrt{j(j+1)}}\, \Tan^{-1} \Bigl( 2\sqrt{j(j+1)} \Bigr) 
\Biggr\}.
\end{equation}

\begin{figure}
\epsfig{file=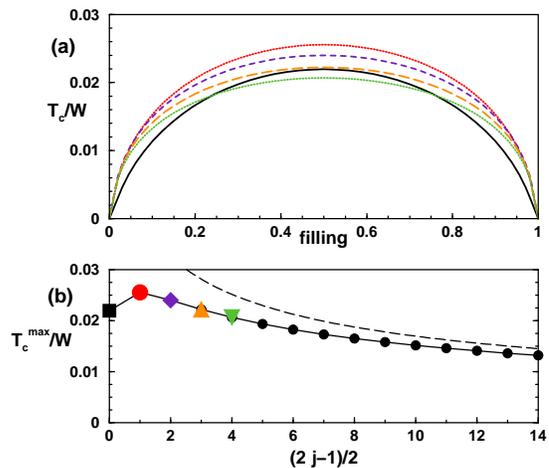, width=0.8\linewidth}
\caption{Ferromagnetic transition temperature for a double-exchange model with general angular 
momentum $j$. (a) The dependence of $\TC /W$ on filling  for $j=1/2$ (solid black), 
3/2 (dotted red), 
5/2 (dashed purple), 7/2 (long-dashed orange), and 9/2 (dotted green).
(b) The maximum transition temperature $\TC^{{\rm max}}/W$ (in a half-filled band 
with $\m =\pm j$) versus $(2j-1)/2$ with the limiting value $1/(4\pi \sqrt{2j+1})$ plotted 
in the dashed curve. Large $j$ values, although unphysical, are
plotted to illustrate the asymptotic behavior.}
\end{figure}

This expression gives a slightly larger value than the exact result obtained from Eq.(\ref{tcs}),
which is used to plot $\TC^{{\rm max}}/W$ versus $(2j-1)/2$ in Fig.~1(b).  
For large $j$, $\TC^{{\rm max}}\approx (W'/4\pi )\{ 1-\pi/4j \}$ saturates at $W'/4\pi $,
as indicated by the dashed line in Fig.~1(b).  
Compared to the maximum $\TC^{{\rm max}} $ for $j=1/2$ of $0.0219W$, $\TC^{{\rm max}}  $ for $j=3/2$ 
of $0.0256W$ is 16\% higher.  Even systems with $j=5/2$ and $7/2$ have higher $\TC $'s
than one with $j=1/2$.  The suppression of $\TC$ for $j=1/2$ is due to the
stronger fluctuations at the smallest angular momentum.

To interpret Fig.~1(b), keep in mind that in the absence of magnetic impurities, a semiconductor is 
characterized by the bandwidth $W$ of the conduction band and by the angular momentum $j$
of the charge carriers.  
So for a class of undoped materials with the same bandwidth, the highest transition 
temperature will be found in the magnetic semiconductor with $j=3/2$.  
A weakness of the present approach is that the local moments with spin $S$
are treated classically whereas the charge carriers with angular momentum $j$ are 
treated quantum-mechanically.  Including the fluctuations of the local moments
will further suppress $\TC $. Therefore, the optimum DMS system will have $S \gg j$ so that 
fluctuations of the local moment can be neglected. This assumption is barely satisfied in Mn-doped 
GaAs, where $S=5/2$ \cite{MacDon05}. 

Having established that the semiconductor with $j=3/2$ has the highest Curie
temperature, we now examine the effect of different masses in the heavy $\m =\pm 3/2$
and light $\m =\pm 1/2$ sub-bands of the parent material.  For GaAs, the band masses are 
$m_h=0.50 m$ and $m_l=0.07 m$ ($m$ is the electron mass) 
with a ratio $r=m_l/m_h = 0.14$ \cite{bla:82}.  As pointed out by Zar\' and 
and Jank\' o \cite{zar:02}, the different band masses introduce magnetic frustration because
the kinetic energy $K=\sum_{\vk, \al \beta }\epsilon_{\vk ,\al \beta }c_{\vk ,\al}^{\dagger }
c_{\vk \beta }$ is only diagonalized when the angular momentum of the charge carriers is quantized along 
the momentum direction (for $r=1$, the quantization axis is arbitrary).  Due to the chirality of 
the holes mediating the ferromagnetism, the moments of a pair of Mn atoms at sites $\vR_1$ and 
$\vR_2$ prefer to align perpendicular to the unit vector $\r $ along the $\vR_2-\vR_1$
direction, as sketched in Fig.~2(a). 

\begin{figure}
\epsfig{file=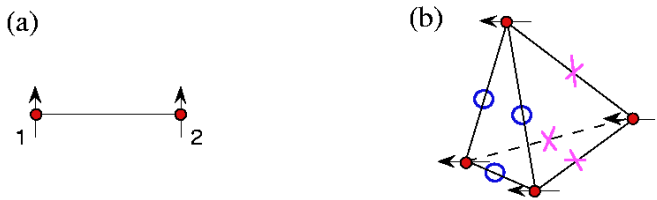, width=\linewidth}
\caption{Cartoon illustrating the magnetic frustration due to the chirality of the holes
mediating the ferromagnetism. (a) The minimum energy configuration for a pair of Mn moments, 
(b) the frustration that
occurs when 4 Mn atoms lie at the corners of a tetrahedron.}
\end{figure}

\begin{figure}
\begin{center}
\begin{minipage}{\linewidth}
\epsfig{file=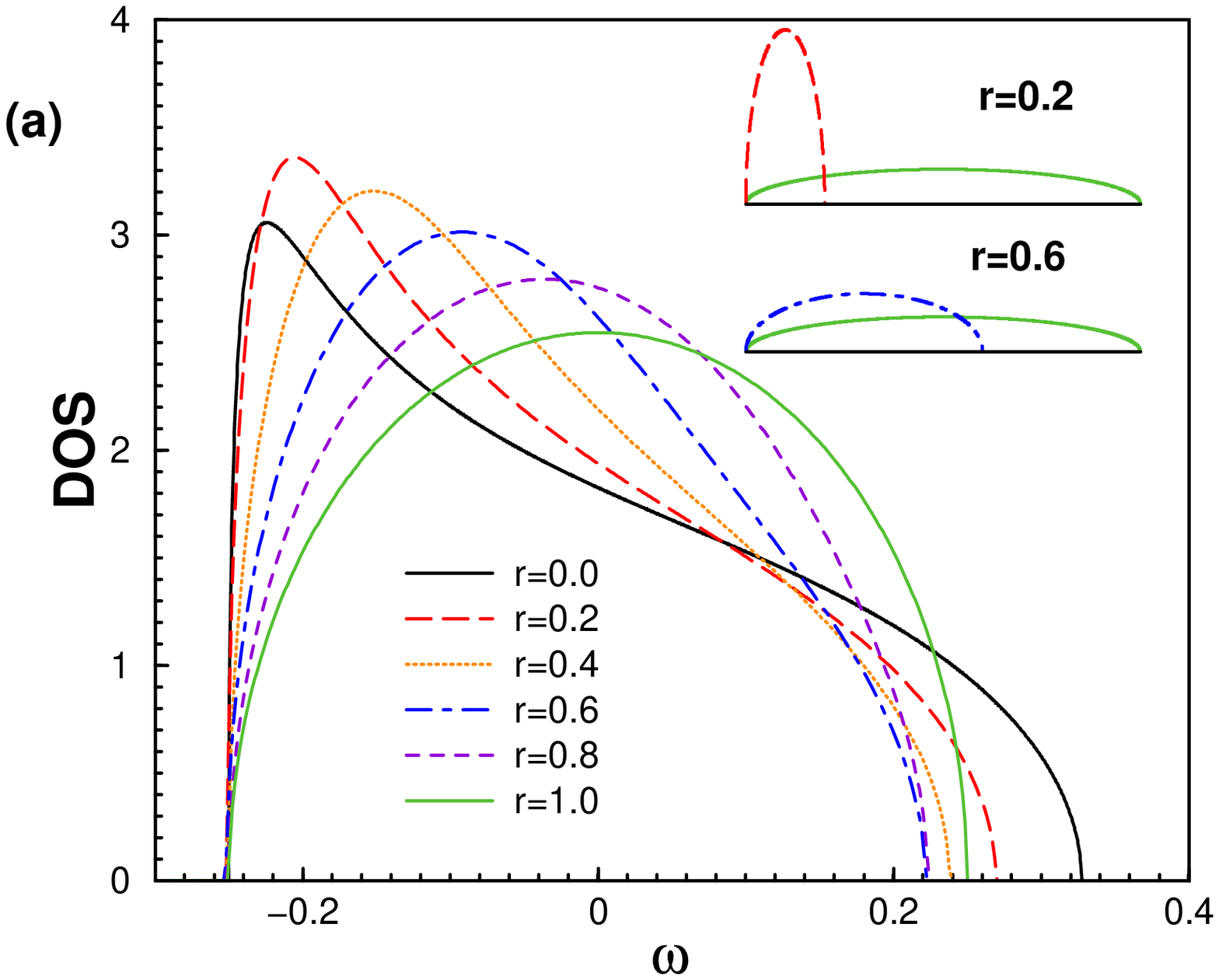, width=0.8\linewidth}
\end{minipage}
\begin{minipage}{\linewidth}
\epsfig{file=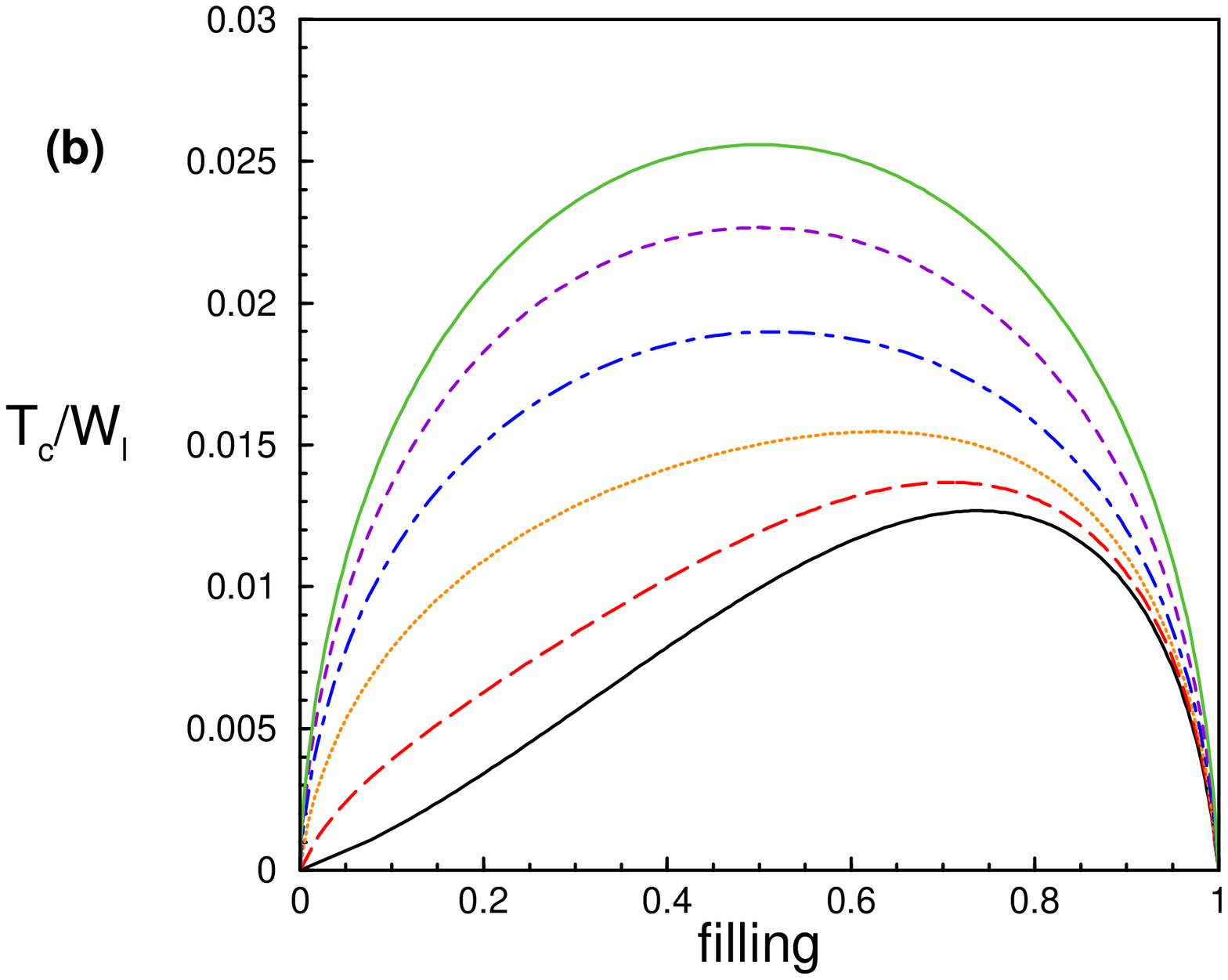, width=0.8\linewidth}
\caption{
Results for a double-exchange model with $j=3/2$ and 
different carrier masses in the heavy and light sub-bands.
(a) The interacting density-of-states for $j=3/2$ 
and a range of ratios r between the effective masses;
at the right hand corner the non-interacting densities-of-states for 
$r=0.2$ and $0.6$. (b) The dependence of $\TC /W_l$ on filling 
for $j=3/2$ and the same values of $r$.}
\end{minipage}
\end{center}
\end{figure}

This anisotropy is easy to understand.  Most of the carriers that couple
Mn atoms 1 and 2 will have momentum ${\bf k}$ along $\r $
and angular momentum ${\bf j}$ parallel to the Mn moments.  If the Mn moments 
lie parallel to $\r $, only heavy holes with angular-momentum component ${\bf j}\cdot \r =3/2$ 
would couple the two Mn.  But if the Mn moments lie perpendicular to $\r $ as in Fig.~2(a), 
holes with angular momentum perpendicular to $\r $ but momentum parallel to $\r $ would contain
all four angular-momentum components ${\bf j}\cdot \r =\pm 3/2$ (heavy) and $\pm 1/2$ (light).
Due to the higher mobility of the light holes, the latter coupling
is more effective.  When four or more Mn atoms are distributed in space, the Mn interactions will 
be frustrated because some of the pairs cannot obtain their lowest energy configuration.  
This is shown schematically in Fig.~2(b), where the circles (crosses) indicate pairs of 
moments that are (not) in their minimum energy configurations. 
The large value of $\TC$ in delta-doped GaAs \cite{Nazmul03} may be related
to the reduced magnetic frustration when Mn ions restricted to two-dimensional planes
can order with their moments perpendicular to the plane.

\begin{figure}[t]
\epsfig{file=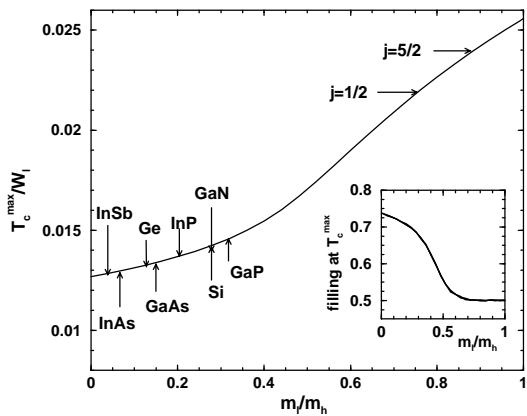, width=0.8\linewidth}
\caption{
Relation between $\TC^{{\rm max}}$ and the ratio of effective masses $r=m_l/m_h$.
In the main panel, $\TC^{{\rm max}} /W_l$  is plotted versus $r=m_l/m_h$, 
with the experimental ratios denoted for several semiconductors.  The horizontal arrows
denote $\TC^{{\rm max}}/W_l$ for $j=1/2$ and $j=5/2$ (equal masses in all sub-bands). 
The inset displays the filling for which the largest Tc is obtained versus $m_l/m_h$.}
\end{figure}

The calculation of $\TC $ is performed using two semi-circular densities-of-states
such as the ones sketched in the right hand corner of Fig.~3(a)
with bandwidths $W_l$ and $W_h=rW_l$.  
As the interaction $\Jc $ is turned on, the interacting density-of-states separates into
four identical sub-bands which are {\it not} particle-hole symmetric.  In Fig.~3(a), we plot 
one of these 
sub-bands for a range of effective mass ratios $r$.  For $r< 1$, the interacting
density-of-states is always weighted towards lower energies, where the heavy holes dominate. 

In Fig.~3(b), we plot $\TC /W_l$ versus filling for the same values
of $r$ used in  Fig.~3(a).  Not surprisingly, $\TC $ decreases as $r$ decreases from 1 
due to the magnetic
frustration introduced by the chirality of the charge carriers.  Hence, the maximum $\TC $ for a 
fixed light bandwidth is obtained when the band masses are identical.  

As $r$ drops from 1 to 0, $\TC^{{\rm max}} /W_l$ plotted in Fig.~4 decreases by 50\% from 0.0256 
to 0.0127.  For the experimental value $r=0.14$ \cite{bla:82}, $\TC^{{\rm max}} \approx 0.0133W_l$ 
has decreased 48\% from its unfrustrated value.  The experimental parameters $r$ for a variety of
semiconductors with $j=3/2$ are also indicated in Fig.~4.  Based on numerical results 
with Mn concentration $x < 1$ \cite{ara:04}, however, the suppression of $\TC $ may not 
be quite so dramatic as suggested by this figure.  Notice also that a system with 
$j=1/2$ or $j=5/2$ and nearly identical band masses may have a higher transition temperature 
than one with $j=3/2$ but very different effective carrier masses.

In the inset to Fig.~4, we plot the filling at which $\TC $ is maximized versus the 
ratio of masses.
This filling varies from $p=0.5$ at $r=1$ to roughly 0.74 at $r=0$.  So for $r < 1$, the 
transition temperature is no longer particle-hole symmetric within the lowest sub-band.  
This may explain why the transition temperature in Mn-doped GaAs continues to 
increase for hole fillings supposedly larger than half of the Mn doping \cite{Ku03,Edmonds02}.
Based on this work, we expect that $\TC $ reaches a maximum for a filling close to 70\%. 

To summarize, we have examined the general dependence of the transition temperature
of a magnetic semiconductor on the angular momentum and filling of the charge carrier
bands.  For a fixed bandwidth of the parent compound, $\TC $ is maximized when $j=3/2$, 
the same angular momentum carried by the holes in GaAs and Ge.  
For $j > 3/2$, the angular momentum of the 
charge carriers effectively suppresses $\TC $ due to the narrowing of the 
impurity band by electronic correlations.  However, the suppression of $\TC $ due
to the different effective masses in $p$-band semiconductors is even more 
significant.
Because of the strong reduction of $\TC $ by the magnetic frustration, 
$s$-band ($j=1/2$) or $d$-band ($j=5/2$) semiconductors
with more nearly equal band-masses may prove to have higher $\TC $'s
than $p$-band ($j=3/2$) semiconductors with disparate effective masses.
For the class of $j=3/2$
semiconductors, $\TC $ will be maximized when the effects of magnetic frustration
are reduced as much as possible. 
This may be achieved either by applying strain, thereby altering the 
band structure, or through digital doping \cite{Nazmul03}.

We gratefully acknowledge useful conversations with Horacio Castillo, Paul Kent
and Igor \v Zuti\' c.
This research was sponsored by the U.S. Department of Energy under contract 
DE-AC05-00OR22725 with Oak Ridge National Laboratory, managed by UT-Battelle, LLC and
by the National Science Foundation under Grant Nos. DMR-0312680 and EPS-0132289 (ND 
EPSCOR).

\end{document}